\documentclass[12pt]{article}
\usepackage{graphicx}
\usepackage{cite}
\usepackage{color}
\newcommand{\bm}[1]{\mbox{\boldmath $#1$}}
\begin{document}
\title{Is the $Y(4260)$ just a coupled-channel signal?}
\author{
Eef van Beveren\\
{\normalsize\it Centro de F\'{\i}sica Te\'{o}rica}\\
{\normalsize\it Departamento de F\'{\i}sica, Universidade de Coimbra}\\
{\normalsize\it P-3004-516 Coimbra, Portugal}\\
{\small http://cft.fis.uc.pt/eef}\\ [.3cm]
\and
George Rupp\\
{\normalsize\it Centro de F\'{\i}sica das Interac\c{c}\~{o}es Fundamentais}\\
{\normalsize\it Instituto Superior T\'{e}cnico, Edif\'{\i}cio Ci\^{e}ncia}\\
{\normalsize\it P-1049-001 Lisboa Codex, Portugal}\\
{\small george@ist.utl.pt}\\ [.3cm]
{\small PACS number(s): 14.40.Gx, 14.40.Lb, 13.25.Gv, 12.39.Pn}\\ [.3cm]
}

\maketitle

\begin{abstract}
The $D_{s}D_{s}^{\ast}$, $D^{\ast}D^{\ast}$, and $D_{s}^{\ast}D_{s}^{\ast}$
$P$-wave channels in the energy region of the $Y(4260)$ charmonium
structure are studied in a coupled-channel model applied to $J^{PC}=1^{--}$
$c\bar{c}$ resonances.  The three channels exhibit enhancements that peak at
4.27 GeV, 4.26 GeV, and 4.33 GeV, respectively, having widths ranging from
80 to 200 MeV. However, no $S$-matrix poles are found, other than those
associated with the $\psi(2D,4160)$ and $\psi(4S,4415)$. The conclusion is
that the observed $Y(4260)$ signal(s) in $\pi\pi J/\psi$ is (are)
probably associated with the opening of the aforementioned channels,
resulting in a resonance-like structure caused by the tail of the
$\psi(3S,4040)$ resonance, roughly midway between the mentioned $P$-wave
thresholds and a sharp kinematical minimum at about 4.4 GeV present in both
the experimental and the model scattering amplitude.
\end{abstract}
\thispagestyle{empty}
\clearpage

The surprising new $J^{PC}=1^{--}$ charm-anticharm $Y(4260)$ enhancement
recently discovered in $\pi^+\pi^-J/\psi$ by the BABAR collaboration
\cite{PRL95p142001}, with mass $\approx$ 4.26 GeV and width $\approx$ 90 MeV,
later confirmed and also seen in $\pi^0\pi^0J/\psi$ as well as $K^+K^-J/\psi$
by the CLEO collaboration \cite{PRL96p162003},
has been studied in a variety of theoretical models \cite{PLB625p212}, namely
as a standard vector charmonium state ($4S$)
\cite{PRD72p031503},
mesonic or baryonic molecule
\cite{PRD72p054023},
gluonic excitation (hybrid)
\cite{PLB631p164},
or $cq\bar{c}\bar{q}$ state
\cite{PRD72p031502}.
In the present paper, we shall present arguments for an again completely
different, yet nonexotic interpretation of the $Y(4260)$.

\begin{figure}[htbp]
\begin{center}
\begin{tabular}{cc}
\includegraphics[height=150pt, angle=0]{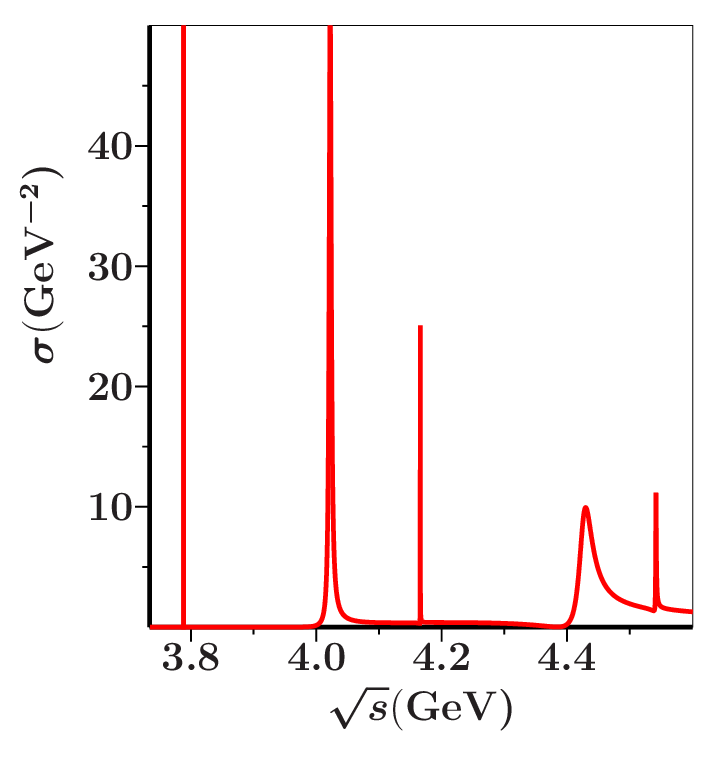}
&
\includegraphics[height=150pt, angle=0]{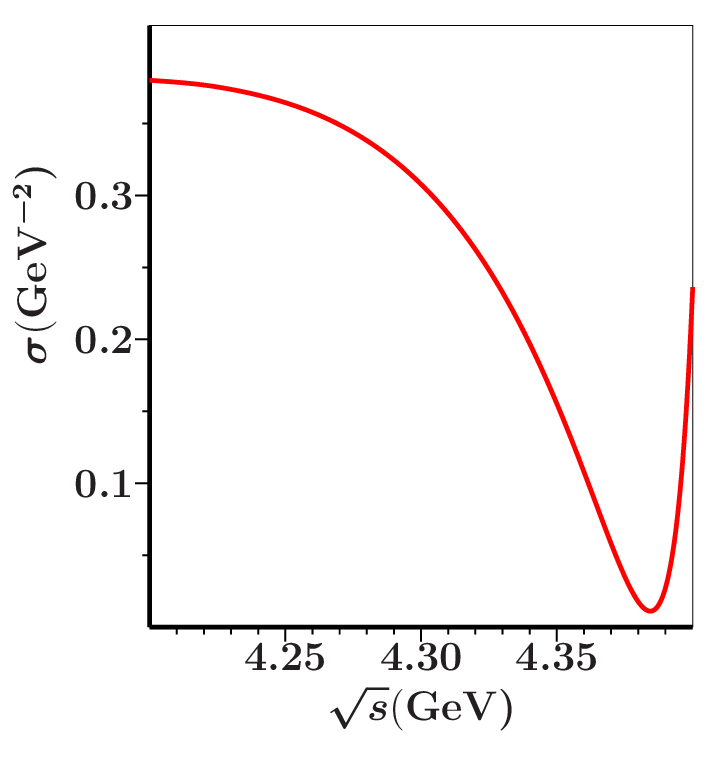} \\[-1cm]
\end{tabular}
\end{center}
\caption[]{The $J^{PC}=1^{--}$ $c\bar{c}$ spectrum of Ref.~\cite{PRD21p772}.}
\label{ccbar}
\end{figure}

In Ref.~\cite{PRD21p772}, a coupled-channel model for quarkonium systems
was presented, which reproduced fairly well the then known charmonium and
bottomonium states, while in Ref.~\cite{PRD27p1527}, using a more realistic
transition potential, also the hadronic widths were reasonably reproduced.
Here, we employ the original model of Ref.~\cite{PRD21p772}, leaving the
parameters unaltered. This yields the spectrum shown in the first figure
in Fig.~\ref{ccbar},
in which one observes the $\psi(1D,3770)$, $\psi(3S,4040)$, $\psi(2D,4160)$,
and $\psi(4S,4415)$, as well as a newly predicted $\psi(3D,4550)$
$J^{PC}=1^{--}$ $c\bar{c}$ state.
For energies in the interval 4.2 GeV to 4.4 GeV, no enhancement is visible
(second figure).
\begin{table}[htbp]
\begin{center}
\begin{tabular}{||rcl|c|c|ll||}
\hline\hline & & & & & & \\ [-10pt]
\multicolumn{3}{||c|}{channel} & threshold & $L_{MM}$ &
\multicolumn{2}{c||}{relative couplings}\\ [3pt]
& & & GeV &
& to $\ell_{c\bar{c}} =0$
& to $\ell_{c\bar{c}} =2$
\\
\hline & & & & & & \\ [-10pt]
$D$ & - & $D$ &   3.73400 & 1
& $     1/    36$
& $     1/   108$
\\ [3pt]
$D_{s}$ & - & $D_{s}$ &   3.93660 & 1
& $     1/    72$
& $     1/   216$
\\ [3pt]
$D$ & - & $D^{\ast}$ &   3.87540 & 1
& $     1/     9$
& $     1/   108$
\\ [3pt]
$D_{s}$ & - & $D^{\ast}_{s}$ &   4.08040 & 1
& $     1/    18$
& $     1/   216$
\\ [3pt]
$D^{\ast}$ & - & $D^{\ast}$ &   4.01680 & 1
& $     7/    36$
& $     1/   270$
\\ [3pt]
$D^{\ast}_{s}$ & - & $D^{\ast}_{s}$ &   4.22420 & 1
& $     7/    72$
& $     1/   540$
\\ [3pt]
$D^{\ast}$ & - & $D^{\ast}$ &   4.01680 & 3
&
& $     7/    60$
\\ [3pt]
$D^{\ast}_{s}$ & - & $D^{\ast}_{s}$ &   4.22420 & 3
&
& $     7/   120$
\\ [3pt]
\hline\hline
\end{tabular}
\end{center}
\caption[]{The various meson-meson channels ($MM$) included in this analysis,
and their relative squared couplings
\cite{ZPC21p291} to $J^{PC}=1^{--}$ $c\bar{c}$ states in $S$ and $D$ wave.}
\label{couplings}
\end{table}
In Table~\ref{couplings}, we summarise the characteristics of the various
meson-meson channels considered in our analysis.

The model of Ref.~\cite{PRD21p772} nonperturbatively accounts for meson loops
below, and meson-meson scattering above threshold. The corresponding
continuum channels contain pairs of $D$, $D_{s}$, $D^{\ast}$ and/or
$D_{s}^{\ast}$ mesons in $P$ or $F$ waves, which are the ones that couple
to vector charmonium. In $DD$, $D_{s}D_{s}$, and $DD^{\ast}$,
as thresholds lie well below the energy interval 4.2--4.4 GeV, we observe no
other structure but the tail of the $\psi(3S,4040)$ resonance
and a sharp kinematical minimum at about 4.4 GeV.

\begin{figure}[htbp]
\begin{center}
\begin{tabular}{ccc}
\includegraphics[height=150pt, angle=0]{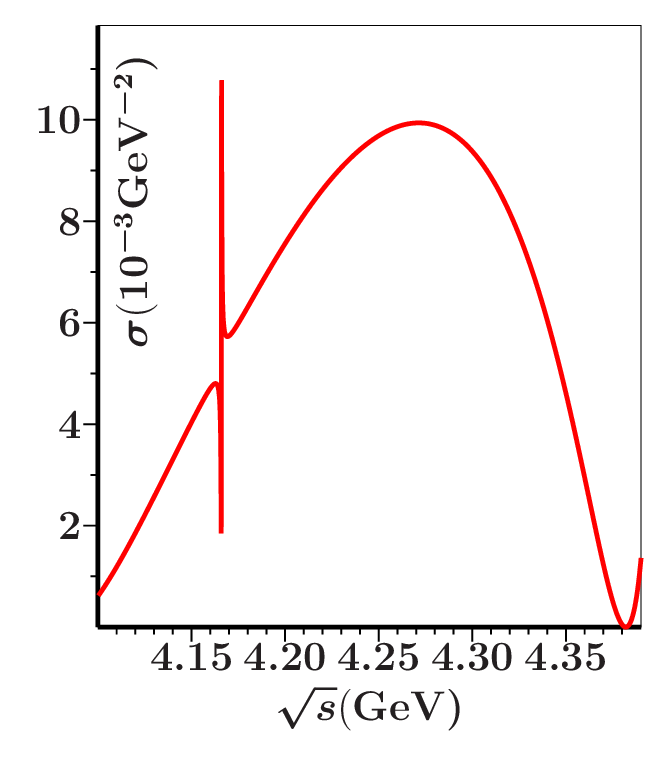}
&
\includegraphics[height=150pt, angle=0]{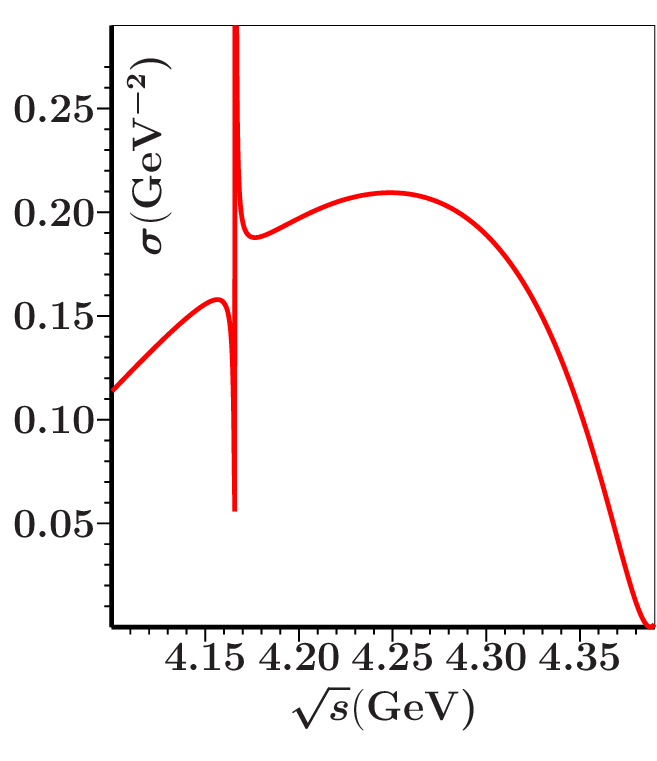}
&
\includegraphics[height=150pt, angle=0]{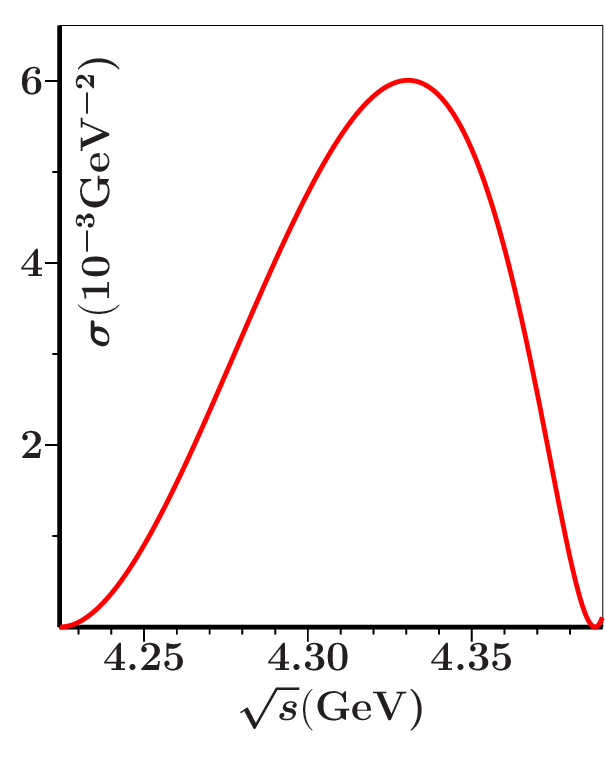}
\\ [-2.5cm]
\hspace*{1cm}\bm{D_{s}D_{s}^{\ast}}
&
\hspace*{1cm}\bm{D^{\ast}D^{\ast}}
&
\hspace*{1cm}\bm{D_{s}^{\ast}D_{s}^{\ast}} \\[1cm] 
\end{tabular}
\end{center}
\caption[]{ $J^{PC}=1^{--}$
model \cite{PRD21p772} $c\bar{c}$ signals in the $D_{s}D_{s}^{\ast}$,
$D^{\ast}D^{\ast}$, and $D_{s}^{\ast}D_{s}^{\ast}$ channels,}
\label{DD}
\end{figure}

On the other hand, the thresholds of $D_{s}D_{s}^{\ast}$, $D^{\ast}D^{\ast}$,
and $D_{s}^{\ast}D_{s}^{\ast}$ lie just below or even inside the latter
energy interval (see Table~\ref{couplings}).
In Fig.~\ref{DD} we show the signals we find in these channels, just above
threshold.
At threshold, the corresponding scattering amplitudes vanish,
because of the relative $P$ waves, starting then to rise with energy.
However, the main two structures that dominate $P$-wave
amplitudes in the energy region 4.1--4.4 GeV
are the $\psi(3S,4040)$ and $\psi(4S,4415)$ resonances, which have large
$S$-wave $c\bar{c}$ components.
Although the amplitudes do not completely vanish because of inelasticities,
the model does produce pronounced dips in all $P$ waves, at slightly different
energies, but all at about 4.4 GeV. This is a rare phenomenon, which can even
be observed quite clearly in the data of Ref.~\cite{PRL95p142001}, thus
deserving further study.
The resulting signal inevitably has a resonance-like shape between the
thresholds and the minimum at $\approx4.4$ GeV. However, no corresponding
resonance pole has been found by us in this energy domain, besides the poles
associated with the $\psi(2D,4160)$ and $\psi(4S,4415)$.

The structure around 4.16 GeV in $D_{s}D_{s}^{\ast}$ and
$D^{\ast}D^{\ast}$ (Fig.~\ref{DD}) is far too narrow in the
model of Ref.~\cite{PRD21p772}, which is an artifact of the
one-delta-shell approximation of the decay mechanism.
According to experiment \cite{PLB592p1}, the $\psi(2D,4160)$ resonance
is 78 MeV wide, hence the ``spikes'' in Fig.~\ref{DD}
appear smeared out over a larger energy interval in reality.
The data shown in FIG.~1 of Ref.~\cite{PRL95p142001} indeed seem
to indicate the presence of precisely such a structure in the
invariant-mass region 4.05--4.21 GeV, where 8 data points behave
exactly as expected for a resonance in the tail of another,
lower-mass resonance, i.e., the $\psi(3S,4040)$. We are well
aware that the authors of Ref.~\cite{PRL95p142001} did \em not
\em \/see this feature in their data. Nevertheless, we are convinced
the $\psi(2D,4160)$ structure is there. In order to support our
\begin{figure}[htbp]
\begin{center}
\begin{tabular}{cc}
\includegraphics[height=150pt, angle=0]{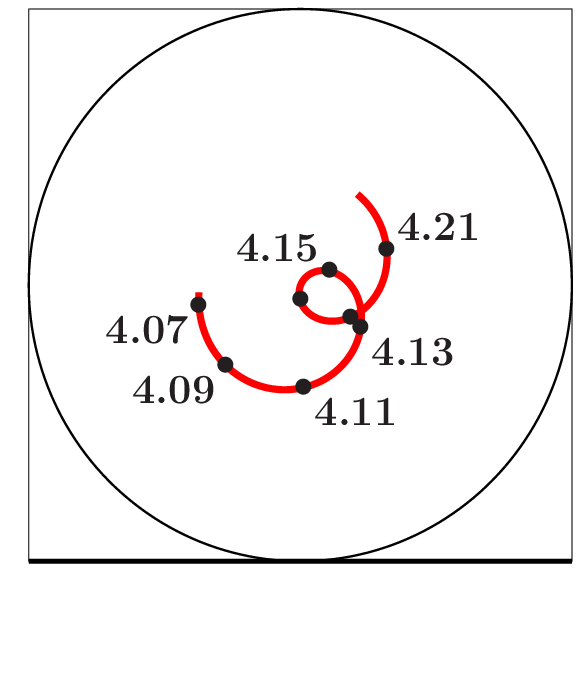}
&
\includegraphics[height=150pt, angle=0]{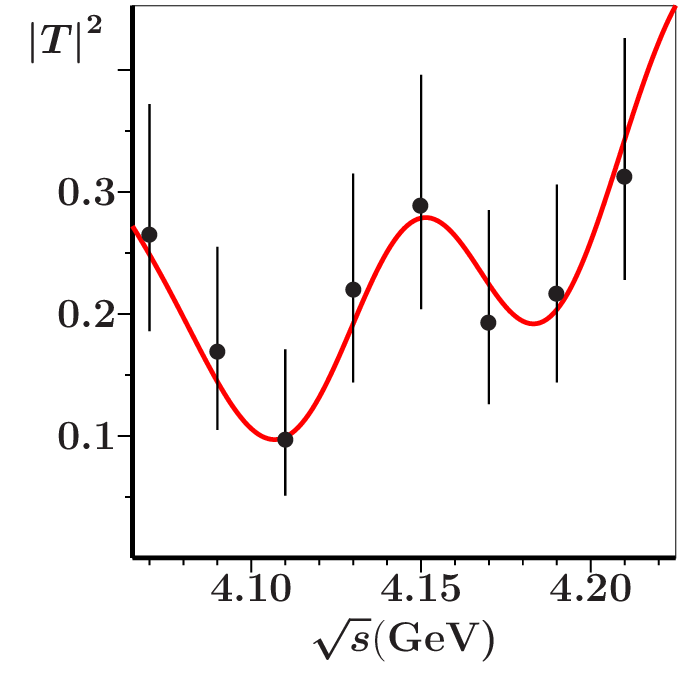} \\[-8mm]
\end{tabular}
\end{center}
\caption[]{Simulated phase motion around the $\psi(2D,4160)$ (left);
corresponding cross section, with 8 data points from FIG.~1 of
Ref.~\cite{PRL95p142001} (right).}
\label{psi2D}
\end{figure}
point,
assuming reasonable values for the amplitude,
we simulate in Fig.~\ref{psi2D} a possible phase motion
that is compatible with the mentioned, and also shown, 8 data
points. We repeat, the depicted  phase motion is just a simulation,
and \em not \em \/a prediction of our model.

Not only does the $\psi(2D,4160)$ come out much too narrow in
the model of Ref.~\cite{PRD21p772}, but actually all resonances are too
narrow.  As a consequence, also the tail of the $\psi(3S,4040)$ 
is in the model somewhat smaller in magnitude than in experiment.
Nevertheless, if we take the results of the model at face value,
we find at $\sqrt{s}=4.25$ GeV a total OZI-allowed hadronic cross section for
decaying vector charmonium of about 0.36 GeV$^{-2}$, which couples with
$\alpha^{2}$ to $e^{+}e^{-}$, resulting in about 7 nbarn.
This is of the correct order of magnitude \cite{PLB592p1}.

The experimental $Y(4260)$ signal seems to be dominantly $f_{0}(980)J/\psi$
(see e.g. Ref.~\cite{PLB625p212}, 5th paper),
which channel opens at about 4.06 GeV, with a maximum at $\approx4.08$ GeV.
Since the $f_{0}(980)$ and $J/\psi$ are in a relative $S$ wave,
with a small $D$-wave component, the amplitude will be maximum at threshold.
This is, of course, the principal reason that the $f_{0}(980)J/\psi$ is
preferred by Nature, as far as non-OZI decays are concerned.
Furthermore, the $f_{0}(980)$ is mostly $s\bar{s}$ \cite{PLB495p300}, hence
it couples preferably to $D_{s}D_{s}^{\ast}$ and $D_{s}^{\ast}D_{s}^{\ast}$.
\begin{figure}[htbp]
\begin{center}
\begin{tabular}{c}
\includegraphics[height=180pt, angle=0]{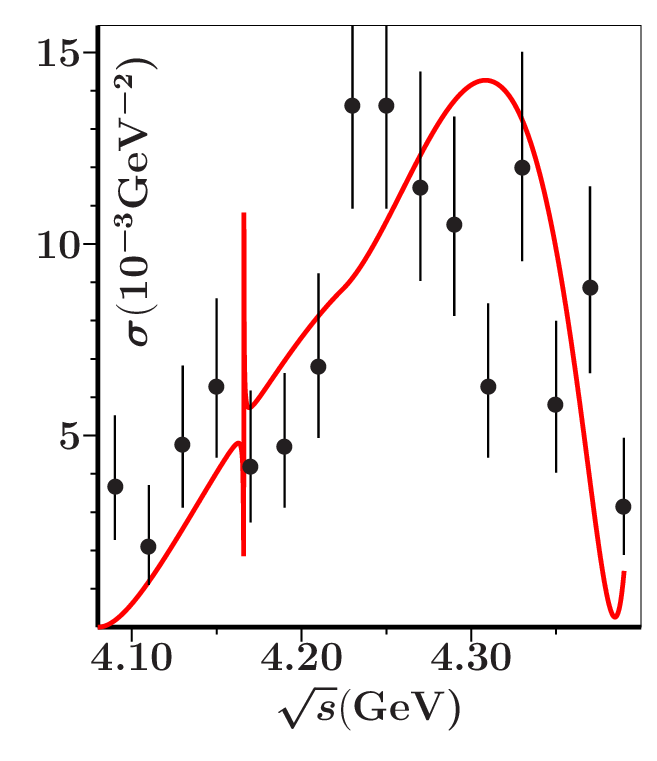} \\[-1cm]
\end{tabular}
\end{center}
\caption[]{ Sum of the $J^{PC}=1^{--}$ $c\bar{c}$ signals in
$D_{s}D_{s}^{\ast}$ and $D_{s}^{\ast}D_{s}^{\ast}$,
for the model of Ref.~\cite{PRD21p772} (with unchanged parameters),
compared to the shape of the data in Ref.~\cite{PRL95p142001}.}
\label{sumss}
\end{figure}
In Fig.~\ref{sumss}, we compare the sum of the signals in these two channels
to the shape of the BABAR data.
If we assume that non-OZI decays are suppressed here so as to
account for only about ten percent of all hadronic decays, then the maximum of
the theoretical curve of Fig.~\ref{sumss} corresponds to 30--40 pb in
$e^{+}e^{-}$, which is in reasonable agreement with the 50 pb estimated in
Ref.~\cite{PRL95p142001}.

In the $^{3\!}P_{0}$-pair-creation picture for unquenching models
of pure confinement, it is assumed that meson pairs are formed
via recombination of the four-quark system (string breaking).
This gives rise to a recombination barrier suppressing non-OZI processes.
However, tunnelling of the $^{3}P_{0}$ pair through the
recombination barrier is not impossible, and many such decay modes
have been observed below the OZI-allowed thresholds.
The $Y(4260)$ enhancement is special in the sense that it is well
above the OZI-allowed thresholds and nevertheless observed in
a OZI-forbidden channel.  Still, it might be seen in the
$D_{s}D_{s}^{\ast}$ and $D_{s}^{\ast}D_{s}^{\ast}$ channels,
and, to a lesser extent, in the $D^{\ast}D^{\ast}$ channel,
all with different line shapes as shown in Fig.~\ref{DD}.
Although technically difficult, the observation of the $Y(4260)$
in these channels could help to sort out its status of a resonance,
which we do not believe it is.

The $e^{+}e^{-}\to\gamma\pi^{+}\pi^{-}J/\psi$ data of BABAR
actually deserve a better analysis than in terms of a simple Breit-Wigner
plus background. As we have shown in the foregoing, the minimum in the
amplitude at about 4.4 GeV is essential for the appearance of a
resonance-like signal. The data contain more such minima.
Some of them may be just statistical fluctuations, others most certainly not.
Now, for spectroscopists, the full structure of the amplitude
is more important than the mere observation of one enhancement.
Nonetheless, we must admit that the excitement about this observation
motivated us to study this energy domain in more detail.
The minimum at 4.4 GeV is a feature of the real data.
Background is not expected to generate such a structure.
Consequently, the real background of the BABAR data must be much smaller
than suggested in the analysis of Ref.~\cite{PRL95p142001}.
More accurate data are probably in production and will be strongly welcomed
by us.

As also confirmed by the CLEO collaboration \cite{HEPEX0604025},
there is no --- or at least no clear --- sign of the vector-charmonium
$S$-state resonances. However, the $D$-state resonances $\psi(1D,3770)$ and
$\psi(2D,4160)$ \em can \em \/be observed in the data of
Ref.~\cite{PRL95p142001}, the former one quite clearly, the latter one also
reasonably well, as we have shown above. This may teach us a new aspect of
non-OZI decay, namely that it mainly couples to the $^{3\!}D_1$ component of
the $c\bar{c}$ system. The presence of the nearby $\psi(2D,4160)$, which
contains a large $D$-state $c\bar{c}$ component in the coupled-channel wave
function, constitutes then a further argument why the observed $Y(4260)$
signal is relatively strong.

Finally, it would also be very helpful if the $\psi(3D,4550)$ predicted in the
model of Ref.~\cite{PRD21p772} could be confirmed by experiment.

\section*{Acknowledgments}
We are indebted to D.~V.~Bugg for enlightening discussions and very
constructive comments. One of us (EvB) wishes to thank
B.~Hiller and A.~Blin for further useful remarks.
This work was supported in part by the {\it Funda\c{c}\~{a}o para a
Ci\^{e}ncia e a Tecnologia} \/of the {\it Minist\'{e}rio da Ci\^{e}ncia,
Tecnologia e Ensino Superior} \/of Portugal, under contract
POCI/FP/63437/2005.

\end{document}